\documentstyle[prl,aps,multicol,mypsfig2]{revtex}

\begin{document}

\title{Methodology for quantum logic gate construction}

\author{%
	Xinlan Zhou$^{1,2}$\thanks{Electronic address: 
				xlz@snow.stanford.edu},
	Debbie W. Leung$^{3,2}$\thanks{Electronic address:
				wcleung@leland.stanford.edu},
	and
	Isaac L. Chuang$^{2}$\thanks{Electronic address: 
				ichuang@almaden.ibm.com}
}

\address{\vspace*{1.2ex}
	$^1$	Department of Applied Physics, Stanford University, 
		Stanford, California 94305-4090}
\address{\vspace*{1.2ex}
	$^2$	IBM Almaden Research Center, 650 Harry Road, 
		San Jose, California 95120 }
\address{\vspace*{1.2ex}
	$^3$	Quantum Entanglement Project, ICORP, JST  \\ 
		Edward Ginzton Laboratory, Stanford University, 
                Stanford, California 94305-4085}

\date{\today}
\maketitle

\begin{abstract}
We present a general method to construct fault-tolerant quantum logic gates
with a simple primitive, which is an analog of quantum teleportation.  The
technique extends previous results based on traditional quantum
teleportation (Gottesman and Chuang, Nature {\bf 402}, 390, 1999)
and leads to straightforward and systematic construction of many
fault-tolerant encoded operations, including the $\pi/8$ and Toffoli
gates. The technique can also be applied to the construction of remote
quantum operations that cannot be directly performed.

\end{abstract}

\section{Introduction}
\label{sec:intro}

Practical realization of quantum information processing requires specific
types of quantum operations that may be difficult to construct.  In
particular, to perform quantum computation robustly in the presence of
noise, one needs fault-tolerant implementation of quantum gates acting on
states that are block-encoded using quantum error correcting
codes~\cite{Shor96,Preskill98,Knill98,Steane99}.
Fault-tolerant quantum gates must prevent propagation of single qubit
errors to multiple qubits within any code block so that small correctable
errors will not grow to exceed the correction capability of the code.  This
requirement greatly restricts the types of unitary operations that can be
performed on the encoded qubits.
Certain fault-tolerant operations can be implemented easily by performing
direct {\em transversal} operations on the encoded qubits, in which each
qubit in a block interacts only with one corresponding qubit, either in
another block or in a specialized ancilla.  Unfortunately, for a given
code, only a few useful operations can be done transversally, and these are
not universal in that they cannot be composed to approximate an arbitrary
quantum circuit.
To obtain a universal set of gates, additional gates have to be constructed
using ancilla states and fault-tolerant measurement.  Although these
additional gates have been constructed
successfully~\cite{Shor96,Knill98,Gottesman98b,Boykin99}, their {\em
ad-hoc} construction is complicated and is not easily generalized.

Another kind of application in which we are challenged to construct useful
quantum operations from a limited set of primitives is in distributed
quantum information processing.  In this problem, certain kinds of
communication between different parties are constrained or prohibited, but
prior distribution of standard states may be allowed.  For example, quantum
teleportation~\cite{Bennett93} demonstrates how an unknown quantum state
can be sent between two parties without sending any quantum information,
using only classical communication and prior entanglement.  Protocols for
distributed state preparation and computation are also
known~\cite{Gottesman98}, but again, they have been largely constructed by
hand and offer neither an explanation of why a particular ancilla state is
required nor a systematic path for generalization.

A general framework for addressing such problems has been presented
in~\cite{Gottesman99}; it uses quantum teleportation as a basic primitive
to enable construction of quantum operations that cannot be directly
performed through unitary operations.  This framework provides systematic
and generalizable construction for an infinite family of fault-tolerant
gates, including the $\pi/8$ and Toffoli gates.  It does not, however, lead
to circuits equivalent to (or as simple as) prior {\em ad-hoc} construction
for the same gates.

In this paper, we provide an extension to the teleportation method of gate
construction with a similar but simpler primitive, which we call ``one-bit
teleportation'' because it uses one qubit instead of two as ancilla.
This method simplifies the construction of~\cite{Gottesman99} and,
furthermore, provides strikingly unified construction of the $\pi/8$,
controlled-phase, and Toffoli gates.  An infinite hierarchy of gates,
including the controlled rotations $\mbox{diag}(1,1,1,e^{i2\pi/2^k})$ used
in the quantum factoring algorithm~\cite{Shor94}, can be constructed with
the present scheme.

The structure of the paper is as follows. First, in
Section~\ref{sec:circuit}, we define one-bit teleportation, and describe
its properties and various guises.  Its application to fault-tolerant gate
construction is presented in Section~\ref{sec:ftqc_gate}, which is followed
in Section~\ref{sec:3examples} with specific circuits for the $\pi/8$,
controlled-phase, and Toffoli gates.  In Section~\ref{sec:remotegate}, we
describe the use of one-bit teleportation to derive the two-bit quantum
teleportation protocol and to construct a remote quantum gate.  We
summarize our results in Section~\ref{sec:concl}.

\section{One-bit teleportation}
\label{sec:circuit}
In standard quantum teleportation, Alice performs a joint measurement of
the unknown qubit and some ancilla, and sends the classical measurement
outcome to Bob, who subsequently reconstructs the unknown state.  No
quantum operation is performed jointly by Alice and Bob, but they need a
certain {\em two}-qubit entangled {\em ancilla} state.  (We refer to any
state other than the original unknown qubit state as ancilla state.)
The same objective, communicating a qubit, can be accomplished in a simpler
manner if Alice and Bob are allowed to perform a quantum gate (such as a
controlled-{\sc not} gate, a {\sc cnot}) between their respective qubits.
In this case, only a {\em single} qubit ancilla in Bob's possession is
required.  We call such a quantum circuit one-bit teleportation, which can
be derived using the following facts:

\begin{itemize}
\item {\bf Fact 1:} An unknown qubit state $|\psi\rangle$ can be swapped with
the state $|0\rangle$ using only two {\sc cnot} gates, as shown in the
following circuit:
\begin{equation}
\parbox{3.25in}{\psfig{file=one-bit-teleport1.epsf,width=1.3in}}
\label{eq:swap}
\end{equation}
Note that in all circuits we show, time proceeds from left to right as is
usual, and conventions are as in~\cite{Nielsen00}.  Throughout this
section, the first and second qubits refer to the registers with respective
initial states $|0\rangle$ and $|\psi\rangle$.

\item {\bf Fact 2:} $X=HZH$, where $X$ and $Z$ are Pauli operators, and
$H$ is the Hadamard gate defined as
\begin{equation}
	H = \frac{1}{\sqrt{2}}
		\left[\begin{array}{cc}
			1 & 1
		\\	1 & -1
		\end{array} \right]
\,.
\end{equation}
Then Eq.~(\ref{eq:swap}) is equivalent to the following circuit:
\begin{equation}
\parbox{3.25in}{\psfig{file=cnot1.epsf,width=1.3in}}
\label{eq:x2}
\end{equation}

\item {\bf Fact 3:} A quantum-controlled gate can be replaced by a
classically-controlled operation when the control qubit is 
measured. 
\begin{equation}
\parbox{3.25in}{\psfig{file=commute-measure.epsf,width=2in}}
\end{equation}
\noindent
The meter represents the measurement of  $Z$, which projects
the measured state onto $|0\rangle$ or $|1\rangle$.  The double line
coming out of the meter carries the 
{\em classical} measurement result,
and $U$ is performed if the measurement result is $|1\rangle$. 
\end{itemize}

In Eq.~(\ref{eq:x2}), the two qubits are disentangled before the second
Hadamard gate.  Therefore, the second qubit can be measured before the
second Hadamard gate without affecting the unknown state in the first
qubit.  Applying fact 3 to Eq.~(\ref{eq:x2}) results in the following
circuit:
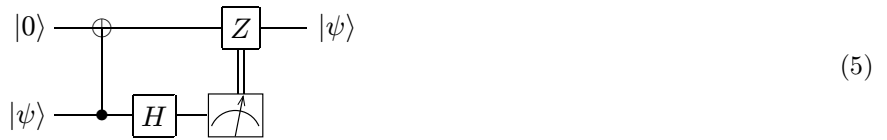
\begin{equation}
\parbox{3.25in}{\psfig{file=one-bit-teleport2.epsf,width=1.9in}}
\label{eq:zteleport}
\end{equation}
The circuit in Eq.~(\ref{eq:zteleport}) uses a {\sc cnot} and only one
qubit for the ancilla.  Therefore it is a one-bit teleportation circuit,
which we refer to as ``$Z$-teleportation'' because a
classically-controlled-$Z$ is applied after the measurement.

Using $Z$-teleportation we can derive other one-bit teleportation
circuits. For instance, the following circuit first teleports the state
$H|\psi\rangle$ using $Z$-teleportation, and then applies $H^{\dagger}=H$
to the teleported state $H|\psi\rangle$ to obtain the original state
$|\psi\rangle$:
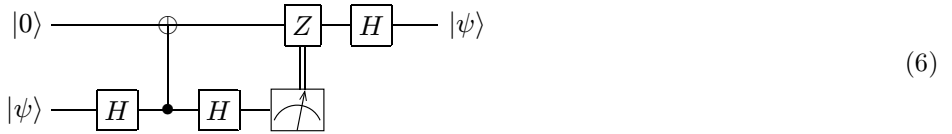
\begin{equation}
\parbox{3.25in}{\psfig{file=one-bit-teleport-x.epsf,width=2.6in}}
\label{eq:ztoxteleport}
\end{equation}
This circuit can be simplified to
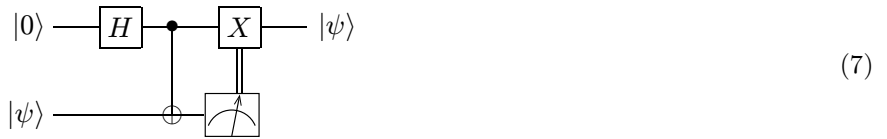
\begin{equation}
\parbox{3.25in}{\psfig{file=one-bit-teleport3.epsf,width=1.9in}}
\label{eq:xteleport}
\end{equation}
which we refer to as ``$X$-teleportation''.
Similarly, we can derive other one-bit teleportation circuits as discussed
in Appendix~\ref{sec:general}.  We will focus on $X$~and $Z$-teleportation
circuits because they are sufficient for our construction in this paper.


$X$~and $Z$-teleportation circuits can both be represented using the same
general structure:
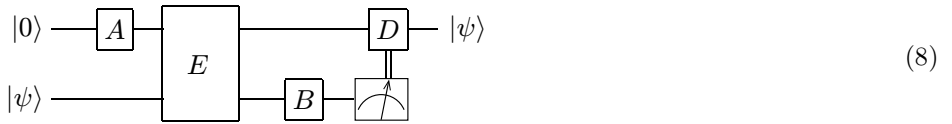
\begin{equation}
\parbox{3.25in}{\psfig{file=xytel.epsf,width=2.6in}}
\label{eq:xztel}
\end{equation} 
where the first qubit (the ancilla qubit) is initially in the $|0\rangle$
state.  For $Z$-teleportation, $A = I$ ($I$ is the $2\times 2$ identity
operator), $B = H, D = Z$, and $E$ is a {\sc cnot} with the first qubit as
its target.  For $X$-teleportation, $A = H, B = I, D = X$, and $E$ is a
{\sc cnot} with the first qubit as its control.

\section{Fault-tolerant gate construction using one-bit teleportation}

\label{sec:ftqc_gate}
In this section, we develop a general method for fault-tolerant gate
construction using one-bit teleportation as a basic primitive.  We will
confine our attention to the Calderbank-Shor-Steane (CSS) codes that are
doubly even and self-dual~\cite{Calderbank96,Steane96,Shor96}, although the
results can be extended to any other stabilizer codes~\cite{Gottesman99}.

\subsection{Fault-tolerant gate hierarchy}
\label{sec:ck}
We first summarize the fault-tolerant gate hierarchy introduced
in~\cite{Gottesman99}. 
Let $C_1$ denote the {\em Pauli group}.  Then for $k\ge 2$, we can
recursively define $C_k$ as
\begin{equation}
	C_k \equiv \{U |UC_1 U^{\dagger} \subseteq C_{k-1}\}
\,.
\label{eq:c3}
\end{equation}
For every $k$, $C_k \supset C_{k-1}$, and 
the set difference $C_k \backslash C_{k-1}$ 
is nonempty. For instance,
$\mbox{diag}(1,e^{i2\pi/2^k}) \in C_k\backslash C_{k-1}$. 

$C_2$ is a group called the {\em Clifford group}~\cite{Gottesman98}, which
is the set of operators that conjugate Pauli operators into Pauli
operators.  Besides the Pauli operators, $C_2$ also contains other
important gates, such as the {\sc cnot}, $H$, and the phase gate $S$
(defined by $S|x\rangle = i^x |x\rangle$ for $x \in \{0,1\}$).  For doubly
even and self-dual CSS codes, any encoded $C_2$ gate has transversal
unitary implementation~\cite{Shor96,Gottesman98b}, which is fault-tolerant.

$C_2$ gates alone, however, are not sufficient for universal quantum
computation~\cite{Gottesman98}. An additional gate outside $C_2$ is
necessary and sufficient to complete universality~\cite{Kitaev00}.  In
particular, adding any one of the following gates in $C_3\backslash C_2$ to
the Clifford group results in a universal set of unitary operations: the
$\pi/8$ gate $T$($T|x\rangle =e^{i\pi x/4}|x\rangle$ for $x\in \{0,1\}$),
the controlled-phase gate $\Lambda_1(S)$ ($\Lambda_1(S)|xy\rangle = i^{x
\cdot y} |xy\rangle$ for $x,y = \{0,1\}$), and the Toffoli gate
(controlled-controlled-{\sc not})~\cite{Shor96,Kitaev97,Boykin99}.

The construction of an encoded operation in $C_3\backslash C_2$ is much
more complicated than that of an encoded operation in $C_2$, and requires
quantum measurement and a particular ancilla state.  But applying a
$C_3\backslash C_2$ gate to certain {\em known} states can be replaced by
direct preparation of the final states, which can be relatively easier as
stated in the following: \\
{\bf Theorem 1:} Let $U$ be an $n$-qubit gate in $C_3$.  Then the encoded
state $U(|0\rangle^{\otimes n})$ can be prepared fault-tolerantly by applying
and measuring $C_2$ operators. \\ {\bf Proof:} See Appendix~\ref{sec:fa}.

Since $C_3$ is closed under multiplication by elements in
$C_2$~\cite{Gottesman00}, Theorem 1 is also applicable when
$|0\rangle^{\otimes n}$ is replaced by $V(|0\rangle^{\otimes n})$ for $V\in
C_2$, because $U |\psi\rangle =UV (|0\rangle^{\otimes n})$ with $UV \in
C_3$.  We will use Theorem 1 in our fault-tolerant logic gate construction.

\subsection{$C_3$ gate construction using one-bit teleportation}
\label{sec:c3}
We now consider a general method of constructing fault-tolerant gates in
$C_3$ using the one-bit teleportation scheme as a primitive.  The basic
idea is the following.  To apply the encoded operation $U$ to an encoded
state $|\psi\rangle$, we can first teleport $|\psi\rangle$ by either $X$~or
$Z$-teleportation, and apply $U$ to the reconstructed $|\psi\rangle$.  The
extra teleportation step can be done fault-tolerantly because both $X$~and
$Z$-teleportation use $C_2$ gates only. It further reduces the problem of
fault-tolerant construction of a quantum logic gate to fault-tolerant
preparation of a particular ancilla state.  The reason for the reduction is
that $U$ is applied to the ancilla, which is originally in the {\em known}
state $|0\rangle$.  If $U$ can be commuted backwards until it is applied to
a {\em known} state without introducing more complicated gates, we can
prepare the resulting known state, without applying $U$ directly, as the
input ancilla.  Using such an ancilla, the reconstructed state after the
{\em modified} one-bit teleportation circuit will be $U|\psi\rangle$.  That
is, the encoded $U$ has been applied to the encoded $|\psi\rangle$
fault-tolerantly.

We now detail the formal construction.
Let $U\in C_3$ be an $n$-qubit gate to be applied to $|\psi\rangle$, an
encoded quantum state with $n$ {\em logical} qubits.
We first teleport each logical qubit using either $X$ or
$Z$-teleportation such that $|\psi\rangle$ is reconstructed in the ancilla,
which is initially in the $|0\rangle^{\otimes n}$ state.
We then apply $U$ to the reconstructed $|\psi\rangle$  to obtain
$U|\psi\rangle$.
This is described by the following quantum circuit:
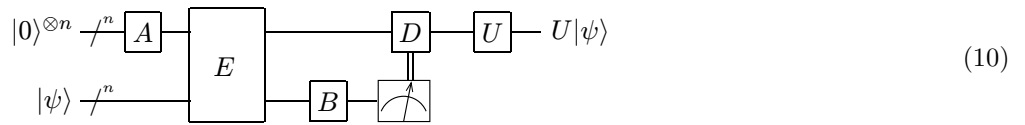
\begin{equation} 
\parbox{3.25in}{\psfig{file=basic1.epsf,width=3.2in}}
\label{eq:basic1}
\end{equation}
In Eq.~(\ref{eq:basic1}), a register (wire) with the symbol ``/$^n$''
represents a bundle of $n$ logical qubits. 
$A$ is a bitwise operation, $A = A_1\otimes \cdots \otimes A_n$, where
$A_i$ acts on the $i^{th}$ logical qubit only.
$B$ is a bitwise operation similar to $A$.
$E$ is a tensor product such that
$E = E_i\otimes \cdots \otimes E_n$, where each $E_i$ is a
{\sc cnot} between the $i^{th}$ logical qubits of $|\psi\rangle$ and the
known ancilla.
The measurement box measures $Z$ bitwise and the double line represents
the $n$-bit classical outcome.  The $i^{th}$ classical bit controls
whether an operator $D_i$ is performed on the $i$$^{th}$ logical state in
the first register.
This is denoted by $D$ for the sake of simplicity.

According to Sec.~\ref{sec:circuit},
if $Z$-teleportation is applied to the $i$$^{th}$ logical qubit, 
$A_i = I, B_i = H, D_i = Z$, and $E_i$ is a {\sc cnot} with the first
qubit as its target;
if $X$-teleportation is applied instead, $A_i = H, B_i = I, D_i = X$, and
$E_i$ is a {\sc cnot} with the first qubit as its control.

We now commute $U$ backwards in time.  
Commuting $U$ with the classically-controlled operation $D$ changes $D$ to
$UDU^{\dagger}$.  As $D \in C_1$ and $U \in C_3$, $UDU^{\dagger} \in
C_2$ can still be performed transversally.
Likewise, commuting $U$ with $E$ changes $E$ to
$UEU^{\dagger}$~\cite{notation}. 
As {\sc cnot} $\notin C_1$, the resulting operation
$UEU^{\dagger}$ may not be in $C_2$ for an arbitrary $U \in C_3$.
To ensure $UEU^{\dagger}\in C_2$,
we only consider $U$ that commutes with $E$ such that
$UEU^{\dagger} = E \in C_2$.
Then Eq.~(\ref{eq:basic1}) becomes
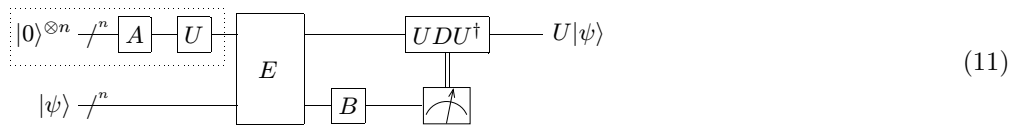
\begin{equation} 
\parbox{3.25in}{\psfig{file=basic3.epsf,width=3.2in}}
\label{eq:basic2}
\end{equation}
All the circuit elements outside the dotted box can be performed
fault-tolerantly.  Therefore, if we can prepare 
the input ancilla in the 
state
$UA(|0\rangle^{\otimes n})$,  we can apply $U\in C_3$ to
any encoded state $|\psi\rangle$ fault-tolerantly.
As $A\in
C_2$, $UA$ is also a $C_3$ operation. 
By Theorem~1, the ancilla state
$UA(|0\rangle^{\otimes n})$ can be created fault-tolerantly.
The stabilizers of such an ancilla state, which will be measured in preparing
the state, can be easily derived.
Recall that when $A_i = I$, $D_i = Z_i$, and when $A_i = H$, $D_i = X_i$.
Therefore, $A_i Z_i A_i^{\dagger} = D_i$ is always true~\cite{coincidence},
and the stabilizers of $UA(|0\rangle^{\otimes n})$ are $U A_i Z_i
A_i^{\dagger}U^\dagger= U D_i U^\dagger \in C_2$.

Using the above method, we can systematically construct interesting
gates in $C_3\backslash C_2$, including the $\pi/8$,
controlled-phase, and Toffoli gate, as will be shown in
Section~\ref{sec:3examples}.

Finally, we remark that the $C_3$ gates commuting with $E$ are not the only
gates that can be performed by the one-bit teleportation scheme. Any $C_3$
gate of the form $U = G_b V G_a$ for $V$ commuting with $E$ and $G_a,G_b
\in C_2$ can be performed using the generalized one-bit teleportation
circuits.  We discuss this in Appendix~\ref{sec:general}.

\subsection{Recursive construction}
\label{sec:recursive}

In this section, we extend our discussion to the gates in $C_k$ and
characterize a class of gates that can be recursively constructed with
one-bit teleportation as a basic primitive.

We prove by induction that the diagonal subset of $C_k$, defined by
$F_k = \{U\in C_k \mbox{ and } U \mbox{ is diagonal}\}$, can be
recursively constructed.

First, when $U \in F_k$, we choose to apply $X$-teleportation to
each logical qubit.  In this case, each $E_i$ is a {\sc cnot} taking the
$i^{th}$ logical qubit in the ancilla as the control bit.  Therefore,
$E$ commutes with $U$ and Eq.~(\ref{eq:basic2}) holds with $A_i = H, B_i =
I$ and $D_i = X_i$ for $i = 1,\ldots,n$. 
Second, since for $U\in F_k$ and $P\in C_1$, $UPU^{\dagger} = \tilde{U}P$
for some $\tilde{U}\in F_{k-1}$~\cite{Gottesman00}, $UD_iU^{\dagger} = UX_i
U^{\dagger} = U_xX_i$ for some $U_x\in F_{k-1}$.  Therefore, if the gates
in $F_{k-1}$ can be performed, the classically-controlled operation
$UD_iU^{\dagger}$ for $U\in F_k$ can also be performed.
Third, the required ancilla $UH^{\otimes n}(|0\rangle^{\otimes n})$ can be
prepared fault-tolerantly with recursive construction as shown in
Appendix~\ref{sec:fa}.
Finally, the gates in $F_2 \subset C_2$ have transversal implementation.
By induction, all the gates in $F_k$ can be performed fault-tolerantly with
recursive application of the one-bit teleportation scheme.

The sets $F_k$ contain many interesting gates, such as $V^{k} =
\mbox{diag}(1,e^{i \pi/ 2^k})$, which are the single qubit $\pi/2^k$
rotations, and $\Lambda_1(V^{k-1}) = \mbox{diag}(1,1,1,e^{i \pi/
2^{k-1}})$, which are the controlled rotations used in the quantum Fourier
transform circuit~\cite{Shor94,Coppersmith94} essential to Shor's factoring
algorithm~\cite{Shor94}.
$F_k$ also includes the multiple-qubit gates $\Lambda_n(V^{l})$ for $n+l\le
k$~\cite{Gottesman00}, where $\Lambda_n(V^{l})$ applies $V^{l}$ to the
$(n+1)^{th}$ qubit if and only if the first $n$ qubits are all in the state
$|1\rangle$.
By the closure property of $F_k$~\cite{Gottesman00}, all products of
$\Lambda_n(V^{l})$ for $n+l\le k$ are in $F_k$.
To perform gates in $F_k$ for $k$ small, the recursive construction we have
described can be more efficient than approximating these gates to the same
accuracy using a universal set of fault-tolerant quantum logic gates.

The gates in $F_k$ are not the only ones that can be constructed using the
one-bit teleportation scheme.
For instance, if $U \in C_k$ is related to an element in $F_k$ by
conjugation with Hadamard gates in the $i_1^{th}$, $\ldots$, $i_l^{th}$
qubits, $E$ can be made to commute with $U$ by applying $Z$ teleportation
to the $i_1^{th}$, $\ldots$, $i_l^{th}$ qubits and $X$-teleportation to the
rest.
The Toffoli gate is an example.
More generally, any gate that is a product of $C_2$ gates and a single
$F_k$ gate can be constructed recursively.

\section{Examples}
\label{sec:3examples}

In this section, we systematically construct three important fault-tolerant
gates in $C_3\backslash C_2$ using the general method described in
Sec.~\ref{sec:ftqc_gate}.  Any one of these gates, together with the
Clifford group, forms a universal set of gates.
For each of the construction, we will derive the required circuit and the
ancilla.  The ancilla can always be prepared fault-tolerantly (see
Appendix~\ref{sec:fa}).

\subsection{The $\pi/8$ gate}

The $\pi/8$ gate, $T$, has the following matrix representation:  
\begin{equation}
	T  = \left[ \begin{array}{cc}
			1 & 0
		\\	0 & e^{i\pi/4}
			\end{array}\right]
\,.
\end{equation}
As $T$ is diagonal, following the recipe in Sec.~\ref{sec:ftqc_gate}, we
choose to apply $X$-teleportation to $|\psi\rangle$ and apply $T$ to the
teleported $|\psi\rangle$:
\begin{equation}
\parbox{3.25in}
{\psfig{file=one-bit-teleport-pi-over-eight.epsf,width=2.4in}}
\end{equation}
We commute $T$ backwards using two facts.  First, $T X T^{\dagger}=
e^{-i\pi/4} S X$, where the phase gate $S$ is defined in Sec.~\ref{sec:ck}.
Second, $T$ commutes with the ${\sc cnot}$ by construction.  Thus, we
obtain a circuit to implement the $\pi/8$ gate (where an irrelevant overall
phase has been ignored):
\begin{equation}
\parbox{3.25in}
{\psfig{file=one-bit-teleport-pi-over-eight-ftqc.epsf,width= 2.4in}}
\end{equation}

All the circuit elements outside the dotted box can be performed
fault-tolerantly.
The dotted box, then,  can be replaced by an ancilla in the state 
\begin{equation}
	|\phi_+\rangle = TH|0\rangle =
	\frac{|0\rangle + e^{i\pi/4}|1\rangle}{\sqrt{2}} 
\,,
\end{equation}
which can be prepared fault-tolerantly as described in
Appendix~\ref{sec:fa}.  Thus, we have derived a circuit and the
corresponding ancilla for performing the fault-tolerant $\pi/8$ gate.  We
note that this re-derives the same circuit and ancilla state used
in~\cite{Boykin99}.

\subsection{The controlled--phase gate}

The controlled-phase gate $\Lambda_1(S)$ (defined in Sec.~\ref{sec:ck}) is
in $C_3$, and forms a universal set of gates~\cite{Kitaev97,Nielsen00}
together with $H$ and {\sc cnot}.

We use the following circuit symbol for $\Lambda_1(S)$:
\begin{equation}
\parbox{3.25in}{\psfig{file=cp.epsf,width=0.7in}}
\end{equation}
$\Lambda_1(S)$ commutes with $Z_i$, and conjugates $X_i$ ($i=1,2$) 
as follows:
\begin{equation}
\parbox{3.25in}{\psfig{file=commute-cp1.epsf,width=2.25in}}
\label{eq:com1}
\end{equation}
\begin{equation}
\parbox{3.25in}{\psfig{file=commute-cp2.epsf,width=2.25in}}
\label{eq:com2}
\end{equation}
where the controlled-$Z$ operation, $\Lambda_1(Z)$, acts on basis states as
$\Lambda_1(Z)|x\rangle|y\rangle = (-1)^{x\cdot y}|x\rangle|y\rangle$.
To construct $\Lambda_1(S)$, we first teleport the two-qubit state
$|\psi\rangle$ and apply $\Lambda_1(S)$.
This linear transformation preserves phase coherence, and thus, it suffices
to consider its action on the basis states $|xy\rangle$.  Since
$\Lambda_1(S)$ is diagonal, we apply $X$-teleportation to both qubit states
such that the {\sc cnot}s in the circuit commute with $\Lambda_1(S)$.
\begin{equation} 
\parbox{3.25in}{\psfig{file=one-bit-teleport-cp.epsf,width=2.6in}}
\end{equation}
Commuting $\Lambda_1(S)$ backwards using the commutation rules in
Eqs.~(\ref{eq:com1})-(\ref{eq:com2}), we obtain a circuit for the
controlled-phase gate:
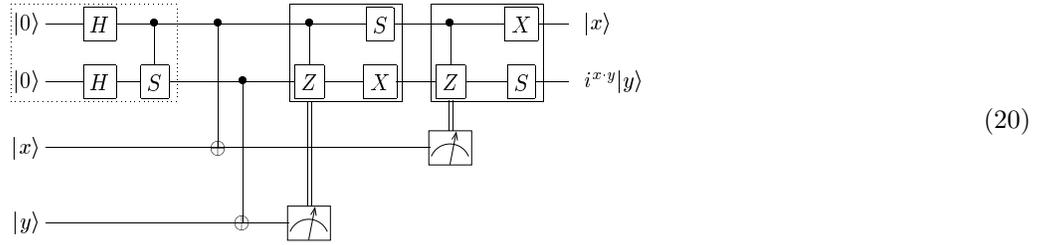
\begin{equation}
\parbox{3.25in}
{\psfig{file=one-bit-teleport-cp-transversal.epsf,width=3.4in}}
\label{eq:cphase}
\end{equation}
where the double lines control all the operations in the corresponding
boxes.
All the circuit elements in Eq.~(\ref{eq:cphase}), except those in the
dotted box, can be performed fault-tolerantly.
Finally, we can replace the dotted box by an input ancilla in the following
state:
\begin{eqnarray} 
	|\phi_+\rangle 
	&=& \Lambda_1(S) (H_1\otimes H_2)|00\rangle 
\\	&=& \frac{1}{2}(|00\rangle + |01\rangle
	+|10\rangle+i|11\rangle)
\,,  
\end{eqnarray}
which can be prepared fault-tolerantly.  This completes the requirement for
performing the controlled-phase gate fault-tolerantly.

\subsection{The Toffoli gate}

To construct the Toffoli gate (controlled-controlled-{\sc not}), we begin
with some useful commutation rules:
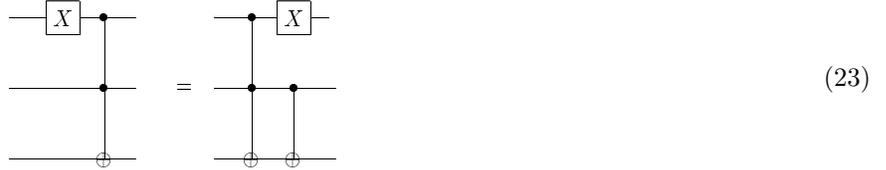
\begin{equation}
\parbox{3.25in}{\psfig{file=commute-toffoli1.epsf,width=1.75in}}
\label{eq:tofcom1}
\end{equation}
\begin{equation}
\parbox{3.25in}
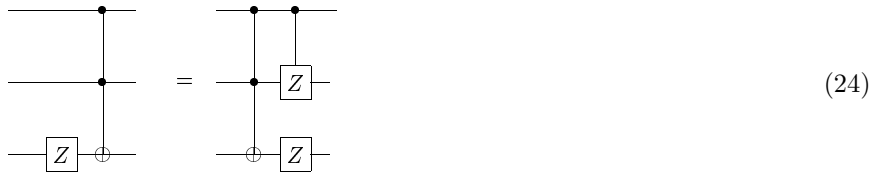
{\psfig{file=commute-toffoli3.epsf,width=1.75in}}
\label{eq:tofcom2}
\end{equation}
As in the controlled-phase gate construction, we demonstrate the
construction on basis states $|xyz\rangle$ for three qubits.  We first
teleport $|xyz\rangle$ and then apply a Toffoli gate.
Since the Toffoli gate is diagonalized by a Hadamard transform on the
target qubit, the choice of $X$-teleportation for the control qubits and
$Z$-teleportation for the target qubit ensures that the three {\sc cnot}s
commute with the Toffoli gate.
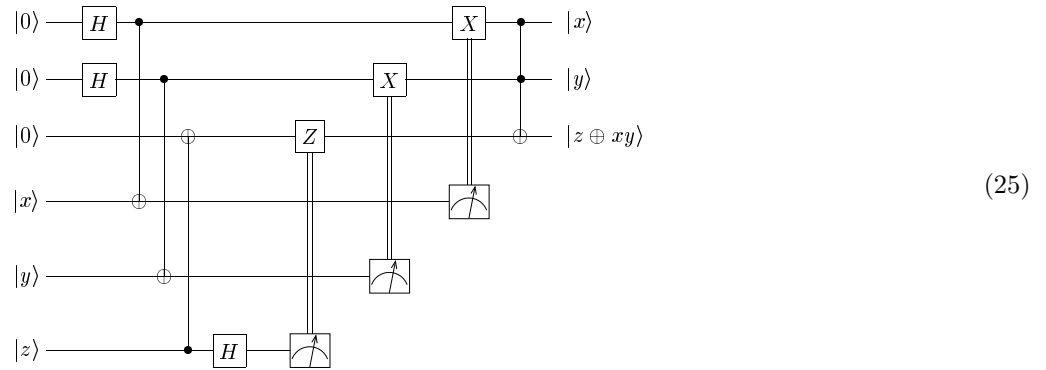
\begin{equation} 
\parbox{3.25in}{\psfig{file=one-bit-teleport-toffoli.epsf,width=3.4in}}
\label{eq:toffoli1}
\end{equation}
Commuting the Toffoli gate backwards to the far left using
Eqs.~(\ref{eq:tofcom1})-(\ref{eq:tofcom2}), Eq.~(\ref{eq:toffoli1}) is
equivalent to
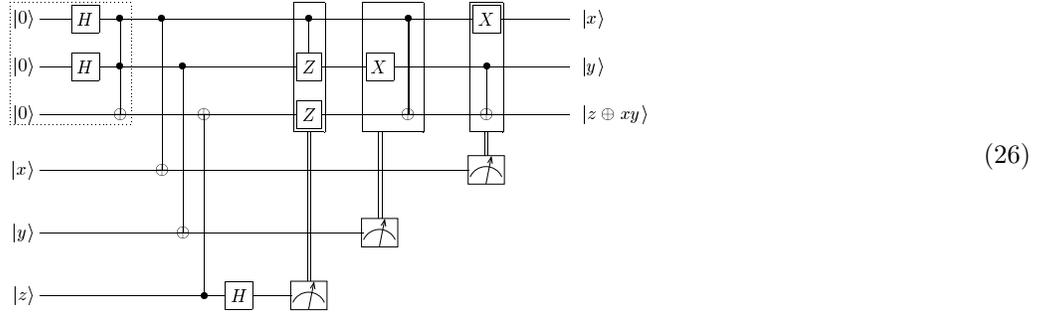
\begin{equation}
\parbox{3.25in}
{\psfig{file=one-bit-teleport-toffoli-transversal.epsf,width=3.4in}}
\label{eq:toffolift}
\end{equation}
All the circuit elements except those in the dotted box can be performed
fault-tolerantly.
It remains to prepare the state created in the dotted box,
\begin{eqnarray}
	|\phi_+\rangle 
	&=& U(H_1\otimes H_2) |000\rangle
\\	&=& \frac{1}{2}(|000\rangle+|010\rangle+|100\rangle+|111\rangle)
\,,
\end{eqnarray}
where $U$ denotes the Toffoli gate. Again this ancilla state can be
prepared fault-tolerantly, as described in Appendix~\ref{sec:fa}.

The ancilla and the quantum circuit derived here are the same as those in
Shor's original construction~\cite{Shor96}.  The one-bit teleportation
scheme elucidates the choice of the ancilla state and the procedure
in~\cite{Shor96}.

\section{Remote gate construction using one-bit teleportation}
\label{sec:remotegate}
The one-bit teleportation scheme, in addition to being useful for
fault-tolerant gate construction, can also be used to design a variety of
remote quantum operations.
Constructing remote quantum operations is related to constructing
fault-tolerant gates in that both require a particular ancilla state to
replace a prohibited operation.
In this section, we use one-bit teleportation as a basic primitive to
derive the quantum circuits and the required ancilla states for the two-bit
quantum teleportation and the remote {\sc cnot}.

\subsection{Two-bit teleportation}

Suppose Alice needs to send a qubit state $|\psi\rangle$ to Bob.  Direct
quantum communication is not allowed, but Alice and Bob can share some
ancilla state.  The question is, how can Alice send $|\psi\rangle$ to Bob?
A well-known solution to this problem is quantum
teleportation~\cite{Bennett93}, which uses an EPR state and classical
communication.
Using one-bit teleportation, we give an alternative derivation of the
required (entangled) ancilla and the required teleportation circuit.

We first construct a circuit to send the unknown state with a prohibited
operation. Then we remove the requirement of such a prohibited operation.
Let $|\psi\rangle$ be the state to be communicated from Alice to Bob. 
Alice can send $|\psi\rangle$ to Bob by applying one-bit teleportation
twice.  Step 1: Alice swaps $|\psi\rangle$ with an ancilla $|0\rangle$
using $X$-teleportation.
Step 2: Alice sends the teleported $|\psi\rangle$ to Bob using
$Z$-teleportation (with a prohibited {\sc cnot} in this step).
The circuit representation for the process is
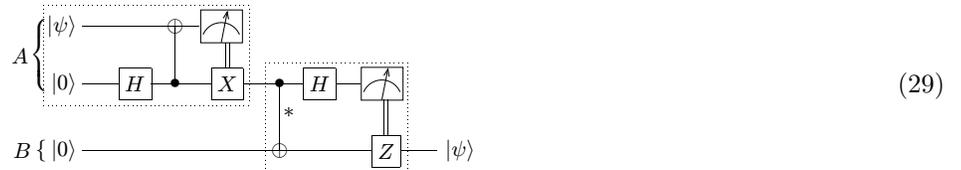
\begin{equation} 
\parbox{3.25in}{\psfig{file=twotel1.epsf,width=2.5in}}
\end{equation}
The prohibited operation ({\sc cnot}), which is marked by an asterisk, can
be commuted backwards using the commutation relation:
\begin{equation}
\parbox{3.25in}{\psfig{file=commute-cnot1.epsf,width=1.9in}}
\end{equation}
This leads to the usual quantum teleportation circuit
\begin{equation}
\parbox{3.25in}{\psfig{file=twotel2.epsf,width=2.5in}}
\label{eq:twotel2}
\end{equation}
In Eq.~(\ref{eq:twotel2}), the prohibited {\sc cnot} acts on the {\em
known} state inside the dotted box, which can be replaced by the following
state it creates:
\begin{equation}
	|\phi\rangle = \Lambda_1(X) H_1 |00\rangle 
	= \frac{1}{\sqrt{2}}(|00\rangle + |11\rangle)
\label{eq:epr}
\end{equation}
In other words, if Alice shares this entangled state with Bob, the state
$|\psi\rangle$ can be sent to Bob without quantum communication.

Note that the classically-controlled-$X$ on the second register only
affects its overall sign, and can be omitted since the second register is
subsequently measured.

An alternative circuit, which accomplishes the same task, can be derived
when $Z$ and $X$-teleportation are used for the two steps instead.  We
start with the following circuit:
\begin{equation}
\parbox{3.25in}{\psfig{file=twotel3.epsf,width=2.5in}}
\end{equation}
Using the commutation rule
\begin{equation}
\parbox{3.25in}{\psfig{file=commute-cnot2.epsf,width=1.9in}}
\end{equation}
we can commute the prohibited {\sc cnot} backwards to obtain an equivalent
quantum teleportation circuit
\begin{equation}
\parbox{3.25in}{\psfig{file=twotel4.epsf,width=2.5in}}
\label{eq:twotel4}
\end{equation}
The disallowed element in the dotted box can be replaced by the EPR state
of Eq.~(\ref{eq:epr}).  The irrelevant classically-controlled-$Z$ on the
second register can be omitted.

The two-bit teleportation circuits of Eqs.~(\ref{eq:twotel2}) and
(\ref{eq:twotel4}) are equivalent to that in~\cite{Brassard98}, but as
mentioned above, they are derived differently.

\subsection{Remote {\sc cnot}}

Suppose Alice and Bob have in their possession quantum states
$|\alpha\rangle$ and $|\beta\rangle$, respectively.
How can they perform a simple distributed computation, a {\sc cnot} from
$|\alpha\rangle$ to $|\beta\rangle$, without communicating any quantum
information between them, but perhaps with the aid of some initially shared
standard quantum state?
A solution to this problem is given in~\cite{Gottesman98}. The {\em ad-hoc}
method employed, however, does not suggest a systematic technique for
deriving the solution, or solutions to generalized versions of this
problem.
Here, we use one-bit teleportation to present a general technique and
derive a different circuit that accomplishes the same task.

Alice and Bob first swap their states with their respective ancilla state
 $|0\rangle$ by one-bit teleportation, and then apply a prohibited {\sc
 cnot}.
The quantum circuit is chosen so that Alice uses $X$-teleportation and Bob
uses $Z$-teleportation:
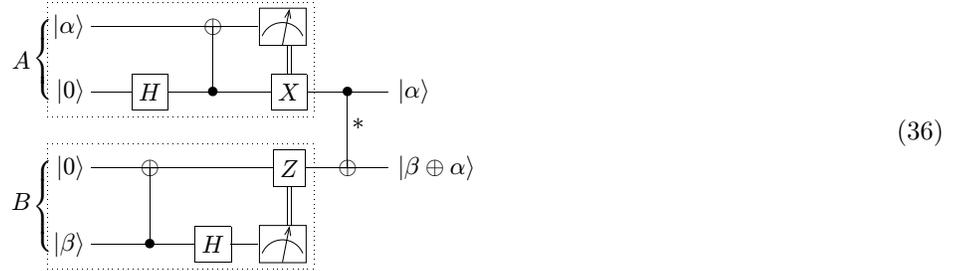
\begin{equation} 
\parbox{3.25in}{\psfig{file=xorb1.epsf,width=2.5in}}
\end{equation}

The prohibited {\sc cnot} can be commuted backwards to obtain a remote
{\sc cnot} circuit:
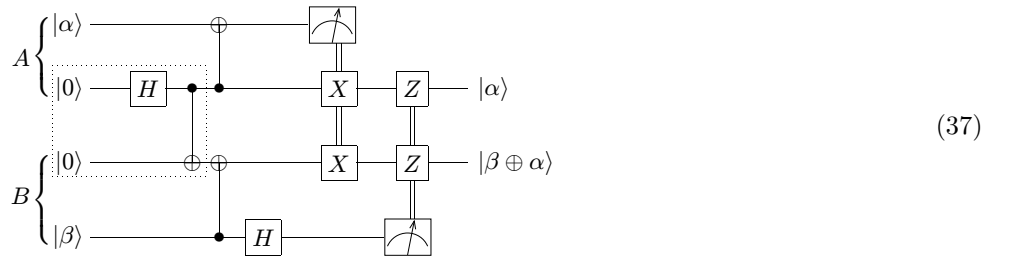
\begin{equation}
\parbox{3.25in}{\psfig{file=xorb2.epsf,width=2.9in}}
\label{eq:remotecn}
\end{equation}
The prohibited operation in the dotted box is applied to a known state, and
can be replaced by the EPR state of Eq.~(\ref{eq:epr}).  Provided such a
shared entangled state is initially available to Alice and Bob, they can
perform a remote {\sc cnot} operation using two bits of classical
communication.

Note that a remote {\sc cnot} can also be constructed by using two-bit
teleportation twice in an obvious way: Bob first sends his qubit
$|\beta\rangle$ to Alice with two-bit teleportation, and then Alice applies
{\sc cnot} to $|\alpha\rangle|\beta\rangle$ and sends the qubit
$|\alpha\oplus \beta\rangle$ to Bob with two-bit teleportation. Such
construction, however, requires two pairs of maximally entangled state and
four bits of classical communication, which is twice that required for the
one-bit teleportation scheme.

Our remote {\sc cnot} construction in Eq.~(\ref{eq:remotecn}) is different
from that in~\cite{Gottesman98}, which can also be derived using the
one-bit teleportation scheme, as described in
Appendix~\ref{sec:gottesmancn}.

Finally, we remark that the two examples of constructing remote operations
strengthen the concept of teleporting quantum logic gates with one-bit
teleportation, as we have shown that if the input ancilla is a special
state related to the {\sc cnot} gate, the reconstructed state is the one to
which a {\sc cnot} gate has been applied.

\section{Conclusion}
\label{sec:concl}
We have presented a systematic technique to construct a variety of quantum
operations, by using a primitive one-bit teleportation scheme.  Such a
scheme reduces the problem of constructing a quantum logic gate to
preparing an ancilla state created by the same gate applied to a known
state.
The usefulness of this technique is particularly manifest for two kinds of
application: fault-tolerant quantum computation and remote quantum
computation, as demonstrated in our construction of the $\pi/8$,
controlled-phase, and Toffoli gates, and the remote-{\sc cnot}. With
recursive application of the one-bit teleportation scheme, we can also
construct an infinite hierarchy of gates fault-tolerantly.

The idea of teleporting quantum logic gates has been used
in~\cite{Gottesman99}, with two-bit teleportation as a primitive, to
perform universal quantum computation.  The two-bit teleportation scheme
allows {\em all} $C_3$ gates to be teleported fault-tolerantly, and {\em
all} $C_k$ gates to be teleported with recursive application of the scheme.
For one-bit teleportation, however, we can only provide {\em sufficient}
conditions for gates in $C_3$ to be teleportable,
namely, any $C_3$ gate that can be written as a product of $C_2$ gates and
a single $C_3$ gate that commutes with {\sc cnot}. It is not known if this
includes all the $C_3$ gates.
The difficulty in describing the exact set of one-bit teleportable $C_3$
gates arises from the requirement for a $C_2\backslash C_1$ gate in the
one-bit teleportation circuit. Such a $C_2\backslash C_1$ gate may be
conjugated outside $C_2$ by a $C_3$ gate, and therefore cannot be directly
performed fault-tolerantly.
This places further constraint on the teleportable $U\in C_k$ for $k>3$.
Because of our present lack of understanding of the general structure and
nature of $C_k$ gates, the distinction between the ultimate capabilities of
the one and two-bit teleportation schemes remains an interesting and
difficult open question.

Nevertheless, as we have shown, one-bit teleportation can provide much
simpler protocols than two-bit teleportation in constructing quantum logic
gates.
This is because one-bit teleportation only requires projective measurement
of $Z$ and as many ancilla qubits as the state to be transformed; two-bit
teleportation, however, requires Bell measurement and twice as many ancilla
qubits as the original state.

At a very general level, the logical gate teleportation schemes reduce the
difficulty of constructing quantum logic gates by using special ancilla
states. This can be useful not only for simplifying hardware requirements,
but also for designing and optimizing computation and communication
protocols~\cite{Cleve97,Cleve97b}.
Even more intriguing, perhaps, is that this result gives us a first glimpse
at what might someday be a standard architecture for a quantum computer: a
simple assembly of one-bit teleportation primitives, capable of universal
quantum computation on quantum data, given the assistance of standard
quantum states that are obtained as commercial resources.
The definition of such a stored-program architecture could be pivotal in
the development of this field,
much as the von~Neumann or Harvard architecture~\cite{Hennessey96} were
important in classical computation.

\section{Acknowledgments}

The relation between fault-tolerant quantum logic gate construction and
teleportation is first alluded to by Shor~\cite{Shor96}.  The $X$ and $Z$
teleportation circuits presented in this paper are due to Charles Bennett
and Daniel Gottesman (unpublished).
We are grateful to Daniel Gottesman for introducing us to the interesting
subject of the $C_k$ hierarchy, and for enlightening discussions.
We thank Professor James Harris and Yoshihisa Yamamoto for support and
encouragement.
This work was supported by the DARPA Ultra-scale Program under the NMRQC
initiative, contract DAAG55-97-1-0341, administered by the Army Research
Office.  D.L. acknowledges support from the IBM Fellowship program and
Nippon Telegraph and Telephone Corporation (NTT).

\appendix

\section{Generalizations of the one-bit teleportation circuits}
\label{sec:general}
The one-bit teleportation circuit used in fault-tolerant gate construction
has three components: a particular input ancilla, a sequence of $C_2$
gates, and finally the measurement and classically-controlled operation.
The teleportability of one-bit teleportation is governed by the sequence of
$C_2$ gates before the measurement.  Using the $X$ and $Z$-teleportation
circuits of Eq.~(\ref{eq:basic2}), any $U\in C_3$ that commutes with $E$
can be teleported.
In this appendix, we derive other one-bit teleportation circuits, which use
different $C_2$ gates, and then discuss their application in constructing
fault-tolerant gates.

By teleporting $G|\psi\rangle$ using $X$-teleportation and applying
$G^{\dagger}$ to the teleported $G|\psi\rangle$, we obtain the following
generalized one-bit telelportation circuit:
\begin{equation}
\parbox{3.25in}{\psfig{file=cliff2.epsf,width=3.0in}}
\label{eq:generalonebit}
\end{equation}
When $G = I^{\otimes n}$ and $H^{\otimes n}$, Eq.~(\ref{eq:generalonebit})
reduces to the $X$ and $Z$-teleportation circuits.

In Sec.~\ref{sec:ftqc_gate}, we showed that all the operations in $F_3$ can
be performed fault-tolerantly using $X$-teleportation.
Here, we generalize the result to show that, if $U\in C_3$ and $U = G_b V
G_a$, where $V\in F_3$ and $G_a,G_b \in C_2$, then $U$ can be performed
fault-tolerantly using the general one-bit teleportation scheme by the
following procedure:

Step 1: Using the circuit of Eq.~(\ref{eq:generalonebit}) with $G=G_a$, we
first teleport the state $|\psi\rangle$ to the ancilla initialized in the
state $|0\rangle^{\otimes n}$, and then apply $U$ to the ancilla.
This can be represented by 
\begin{equation} 
\parbox{3.25in}{\psfig{file=cliff3.epsf,width=3.4in}}
\end{equation}

Step 2: Commuting $U$ backwards, one obtains 
\begin{equation}
\parbox{3.25in}{\psfig{file=cliff4.epsf,width=3.4in}}
\label{eq:s2}
\end{equation}
Note that the new classically-controlled operation is $G_b V X^{\otimes n}
V^{\dagger} G_b^{\dagger}$, which is in $C_2$ because $V X^{\otimes n}
V^{\dagger} \in C_2 $.  Therefore, all the circuit elements can be
performed fault-tolerantly, except those in the dotted box, which can be
replaced by an ancilla in the state $V H^{\otimes n}(|0\rangle^{\otimes
n})$.

There are $C_3$ gates that cannot be constructed using $X$ and
$Z$-teleportation {\em directly}, but can be constructed using other
one-bit teleportation circuits.  For instance, the controlled-Hadamard gate
$\Lambda_1(H_2)\in C_3\backslash C_2$ does not commute with $E$ in
Eq.~(\ref{eq:basic1}) for all possible combinations of $X$ and
$Z$-teleportation circuits, but $\Lambda_1(H_2)$ can be written as
$G_bVG_a$ with $G_a = Q^{\dagger}_2$, $G_b = \Lambda_1(X_2) Q_2$ and $V =
T_1 \Lambda_1(S_2^{\dagger})$, where $Q = S^{\dagger}HS\in C_2$. Thus,
$\Lambda_1(H_2)$ can still be performed using the general one-bit
teleportation scheme.

We remark that a $C_3$ gate $U=G_bVG_a$ with $G_a,G_b\in C_2$ and $V\in
F_3$ can be performed {\em indirectly} by applying $G_a,V$ and $G_b$ in
sequence, where $V$ is applied by $X$-teleportation.
If the operations in the generalized one-bit teleportation circuit, $G_b,
\Lambda_1(X)$, and $G_a$ of Eq.~(\ref{eq:s2}), are also considered, the
total requirements to perform $U$ by such indirect implementation and by
direct one-bit teleportation are almost the same.
But if we are given different one-bit teleportation circuits as primitives,
we can use them to directly teleport different sets of $C_3$ gates. In
other words, if we are given the circuit of Eq.~(\ref{eq:basic2}), using an
input ancilla in the state $UA(|0\rangle^{\otimes n})$, we can teleport
$U\in C_3$ that commutes with $E$; if we are given the circuit of
Eq.~(\ref{eq:s2}), using an input ancilla in the state $VH^{\otimes
n}(|0\rangle^{\otimes n})$, we can teleport $U$ in the form of $G_b V G_a$.
In this sense, then, the generalized one-bit teleportation circuits are
interesting and allow more gates in $C_3$ to be teleported directly.

\section{Fault-tolerant state preparation}
\label{sec:fa}
In this section, we first prove Theorem 1 in Sec.~\ref{sec:ck} by
construction.  We then show how to create the three ancilla states in
Sec.~\ref{sec:3examples} fault-tolerantly.  Finally, we explain how to
prepare a class of encoded quantum states fault-tolerantly by recursive
application of the one-bit teleportation scheme.

\subsection{Fault-tolerant preparation of quantum states}
A stabilizer of a quantum state is a quantum operator that transforms the
state to itself.  Let ${\cal C}$ be the codeword space corresponding to an
$[[m,n]]$ stabilizer code, which encodes $n$ logical qubits using $m$
physical qubits.  The stabilizer $S$ of ${\cal C}$ is an Abelian subgroup
of the Pauli group, or $C_1$, such that $|\psi\rangle\in {\cal C}$ if and
only if $\forall M\in S, M|\psi\rangle = |\psi\rangle$.
By performing error correction for the stabilizer code, we can project an
arbitrary state onto an encoded state in ${\cal C}$
~\cite{Gottesman98,Preskill98}.

The stabilizer $S$ has $2^{m-n}$ elements generated by $m-n$ independent
operators in $C_1$, and defines a quantum code of dimension $2^n$.  Each
encoded state is, then, determined by $n$ extra independent
stabilizers. (In the following, we will restrict our discussion to the
codeword space and exclude the stabilizers of the code from the stabilizers
of an encoded state.)  For instance, the encoded $|0\rangle^{\otimes n}$ is
determined by $Z_i$ for $i=1,\ldots,n$, where $Z_i$ is the encoded $Z$ on
each logic qubit.  In general, stabilizers need not commute with one
another and need not square to the identity.
But an independent set of stabilizers can always be chosen to be a mutually
commuting set of elements that square to the identity.  This is because
$|\psi\rangle = U(|0\rangle^{\otimes n})$ for some encoded $U$, leading to
a possible choice $\mbox{St}(|\psi\rangle) \equiv \{UZ_iU^{\dagger}, i =
1,\ldots, n\}$ with the desired properties.  We restate the above as a
lemma: \\
{\bf Lemma 1:} For any $|\psi\rangle$, $\mbox{St}(|\psi\rangle)$ can be
chosen such that $\forall M,N\in \mbox{St}(|\psi\rangle)$, (a) $M^2 = I$
and (b) $[M,N] = MN-NM=0$. \\
Note that the elements in $\mbox{St}(|\psi\rangle)$ are all valid encoded
operations, and their actions preserve the codeword space.

As a quantum state is the simultaneous $+1$ eigenstate of its stabilizers,
the state can be prepared by projecting an arbitrary encoded state onto the
simultaneous $+1$ eigenstate of its stabilizers.
In the following we will show how to create a class of quantum states
fault-tolerantly by measuring their stabilizers.

Given a quantum state $|\psi\rangle$ encoded with an $[[m,n]]$
stabilizer code, the operator $M\in C_2$ with $M^2=I$ can be measured
fault-tolerantly on $|\psi\rangle$ as follows.  First, we prepare a cat
state
\begin{equation}
	|\mbox{cat}\rangle \equiv \frac{1}{\sqrt{2}}
	(|\bar{0}\rangle + |\bar{1}\rangle)
\,,
\end{equation}
where $|\bar{i}\rangle$ consists of $m$ physical qubits in the state
$|i\rangle$ ($i = 0,1$).  (The cat state cannot be created
fault-tolerantly, but it can always be verified~\cite{Shor96}.)
For the doubly even and self-dual CSS codes, the encoded $M\in C_2$ can be
written as $M=M^1\otimes \cdots \otimes M^m$, where $M^j$ acts only on the
$j^{th}$ physical qubit of each block of the encoded state $|\psi\rangle$.
For each $j$, we perform controlled-$M^j$ with the $j^{th}$ qubit of the
cat state as the control bit and the $j^{th}$ qubit of $|\psi\rangle$ as
the target qubit. Effectively, a cat-state-controlled-$M$ is applied to the
state $|\mbox{cat}\rangle|\psi\rangle$ with transversal operations leading
to the state
\begin{eqnarray}
	& &\frac{1}{\sqrt{2}}|\bar{0}\rangle |\psi\rangle + \frac{1}{\sqrt{2}}
	|\bar{1}\rangle M|\psi\rangle
\\	&=& \frac{1}{2} (|\bar{0}\rangle + |\bar{1}\rangle) (I+M) |\psi\rangle
	+ \frac{1}{2} (|\bar{0}\rangle - |\bar{1}\rangle) (I-M) |\psi\rangle
\,.
\end{eqnarray}
Note that as $M^2 = I$, 
$(I\pm M) |\psi\rangle$ are $\pm 1$ eigenstates of $M$ for any
$|\psi\rangle$. 

We can measure the cat state fault-tolerantly using the procedure described
in~\cite{Shor96,Preskill98} to distinguish $|\bar{0}\rangle +
|\bar{1}\rangle$ from $|\bar{0}\rangle - |\bar{1}\rangle$. (We omit the
unimportant normalization factors.)  If we obtain $|\bar{0}\rangle +
|\bar{1}\rangle$, the encoded state is projected onto $(I+M)|\psi\rangle$,
the $+1$ eigenstate of $M$; otherwise the resulting encoded state is
$(I-M)|\psi\rangle$, the $-1$ eigenstate of $M$, which may be transformed
to a $+1$ eigenstate of $M$ by the following Lemma.\\
{\bf Lemma 2:} If $M\in C_2, M^2 = I$, and there exists $Q\in C_2$ such
that $\{M,Q\} = MQ+QM = 0$, then we can always transform an arbitrary
encoded state $|\psi\rangle$ onto a $+1$ eigenstate of $M$ using
fault-tolerant operations. The resulting $+1$ eigenstate is either
$(I+M)|\psi\rangle$ or $(I+M)Q|\psi\rangle$, which can be written jointly
as $(I+M)Q^{a}|\psi\rangle$ for $a = 0$ or $1$. \\
{\em Proof:} We have shown that we can project an arbitrary encoded state
$|\psi\rangle$ onto $(I\pm M)|\psi\rangle$, the $\pm 1$ eigenstate of
$M$. If the resulting state is $(I+M)|\psi\rangle$, we are done; otherwise,
we apply $Q\in C_2$ fault-tolerantly to $(I-M)|\psi\rangle$.
Since $Q$ anticommutes with $M$, it transforms the $-1$ eigenstate of $M$
to a $+1$ eigenstate of $M$ as follows: $Q(I-M)|\psi\rangle =
(I+M)(Q|\psi\rangle)$.  Thus, we can always obtain a $+1$ eigenstate of
$M$, which is $(I+M)Q^{a}|\psi\rangle$ for $a = 0$ or $1$.

Next we will show that a special class of quantum states can be
created fault-tolerantly. \\
{\bf Lemma 3:} If $\mbox{St}(|\psi\rangle) = \{M_1,\ldots,M_n\} \subset
C_2$ and $\forall M_i \in \mbox{St}(|\psi\rangle)$ there exists $Q_i\in
C_2$ such that $\{M_i,Q_i\} = 0$, and $[M_i,Q_j]= 0$ for $i\neq j$, then
$|\psi\rangle$ can be created fault-tolerantly by measuring the elements
in $\mbox{St}(|\psi\rangle)$ fault-tolerantly.\\
{\em Proof:} By Lemma 1, $\forall i$, $M_i^2 = I$. Starting from {\em any}
encoded state $|\phi\rangle$, we measure $M_1,\ldots,M_n$ sequentially, and
after each measurement we apply the corresponding operation $Q_i$ if the
projected state is the $-1$ eigenstate of $M_i$.  By Lemma 2, the resulting
state is
\begin{eqnarray}
	|\psi\rangle 
	&=& (I+M_n)Q_n^{a_n}\cdots (I+M_1)Q_1^{a_1}|\phi\rangle
\\	&=& (I+M_n)\cdots (I+M_1) Q_n^{a_n}\cdots Q_1^{a_1}|\phi\rangle
\,,
\end{eqnarray}
where $a_i = 0$ or $1$, and we have used the fact that $[M_i,Q_j]=0$ for
$i\neq j$.  As $[M_i,M_j] = 0$, it is easily verified that $\forall i, M_i
|\psi\rangle = |\psi\rangle$, and $|\psi\rangle$ is the desired state that
has been created fault-tolerantly.

Theorem 1 in Sec~\ref{sec:ck} immediately follows:\\
{\bf Theorem 1:} $\forall U\in C_3$, 
$U$ can be applied to the encoded $|0\rangle^{\otimes n}$ state using $C_2$
operators and fault-tolerant measurement of $C_2$ operators. \\
{\em Proof:} Applying $U\in C_3$ to the encoded
$|0\rangle^{\otimes n}$ state is equivalent to preparing the state 
$|\psi\rangle = U(|0\rangle^{\otimes n})$, which has stabilizers
$M_i = UZ_iU^{\dagger}$ for $i = 1,\ldots,n$.
Define $Q_i\equiv UX_iU^{\dagger}\in C_2$ for each $i$. 
Then $\{Z_i,X_i\} = 0$ implies $\{M_i,Q_i\} = U\{Z_i, X_i\}U^{\dagger} =
0$, and for $i\neq j$, $[Z_i,X_j] = 0$ implies $[M_i,Q_j] = U[Z_i,
X_j]U^{\dagger} =0$. Thus by Lemma 3, the state $|\psi\rangle$ can be
created fault-tolerantly.

\subsection{Examples}
To prepare a specific encoded state from an {\em unknown} encoded state we
need to measure {\em all} its independent stabilizers.
When the initial state is a {\em known} encoded state related to the
desired state, we may not have to measure all the independent stabilizers.
For instance, given two encoded states $|\phi\rangle$ and $|\phi'\rangle$
with $\mbox{St}(|\phi\rangle) = \{M_1,\ldots, M_k,M_{k+1},\ldots, M_n\}$
and $\mbox{St}(|\phi'\rangle) = \{M_1,\ldots,M_k,M_{k+1}',\ldots,M_n'\}$,
the following state
\begin{equation}
	(I+M_n)\cdots (I+M_{k+1}) Q_n^{a_n} \cdots
	Q_{k+1}^{a_{k+1}} |\phi'\rangle
\end{equation}
is the simultaneous $+1$ eigenstate of $M_i$ for $i = 1,\ldots,n$.  Thus,
starting from $|\phi'\rangle$, we can prepare the encoded state
$|\phi\rangle$ by measuring only the $n-k$ different stabilizers.
In the following, we will construct an initial state, with which, the
desired state can be obtained by measuring only a single stabilizer.

Assume we want to prepare the encoded state $|\psi_+\rangle =
U(|0\rangle^{\otimes n})$ for $U\in C_3$. Define $M_i$ and $Q_i$ for $i =
1,\ldots, n$ as in the proof of Theorem 1.
Then $Q_i|\psi_+\rangle$ is a $-1$ eigenstate of $M_i$ such that $\langle
\psi_+|Q_i|\psi_+\rangle = 0$, and
the following state
\begin{equation}
	|\psi\rangle =
	\frac{1}{\sqrt{2}}(|\psi_+\rangle+Q_i|\psi_+\rangle) 
\end{equation} 
is different from $|\psi_+\rangle$ by only one independent stabilizer:
$Q_i$ has replaced $M_i$. Therefore, the state $|\psi\rangle$ also
satisfies the conditions of Lemma 3, and can be prepared fault-tolerantly.
It follows that to obtain the state $|\psi_+\rangle$, we only need to
measure the single stabilizer $M_i$ on $|\psi\rangle$.

To prepare an encoded state $|\psi_+\rangle$ by preparing $|\psi\rangle$
first can be simpler than directly preparing $|\psi_+\rangle$ from an
arbitrary encoded state if $|\psi\rangle$ itself can be prepared
easily. For instance, when $|\psi\rangle$ is a product state, it can be
prepared by measuring only single qubit operators.  We will describe how to
prepare the required ancilla states for the three gates in
Sec.~\ref{sec:3examples}.  When the required ancilla $|\psi_+\rangle$ is an
entangled state with multiple-qubit stabilizers, we will construct it by
preparing an easier state $|\psi\rangle$ first.

\subsubsection{Fault-tolerant preparation of the ancilla required for $T$ gate}
The required ancilla for constructing the $\pi/8$ gate, $T$,   is 
\begin{equation}
	|\psi_+\rangle = TH|0\rangle =
	\frac{|0\rangle + e^{i\pi/4}|1\rangle}{\sqrt{2}} 
\,,
\end{equation}
with stabilizer
\begin{equation}
	M = (TH)Z(TH)^{\dagger} = e^{-i\pi/4} S X
\,,
\end{equation}
which anticommutes with $(TH)X(TH)^{\dagger} = Z$.  Then starting from any
encoded state, we can measure $M$, and apply $Z$ if the projected state is
the $-1$ eigenstate, to create the state $|\psi_+\rangle$ fault-tolerantly.

\subsubsection{Fault-tolerant preparation of the ancilla required for
controlled-phase gate}
The required ancilla for constructing the controlled phase gate is
\begin{eqnarray} 
	|\psi_+\rangle 
	&=& \Lambda_1(S) (H_1\otimes H_2)|00\rangle 
\\	&=& \frac{1}{2}(|00\rangle + |01\rangle
	+|10\rangle+i|11\rangle)
\,, 
\end{eqnarray}
with stabilizers $M_i = \Lambda_1(S) (H_1\otimes H_2) Z_i (H_1\otimes H_2)
\Lambda_1(S^{\dagger})$ for $i = 1,2$. Using
Eqs.~(\ref{eq:com1})-(\ref{eq:com2}), 
\begin{eqnarray}
	M_1 &=& (X_1\otimes S_2) \Lambda(Z)
\,,
\\	M_2 &=& (S_1\otimes X_2) \Lambda(Z)
\,.
\end{eqnarray}
The corresponding operator that anticommutes with $M_i$ is $Q_i =
\Lambda_1(S) (H_1\otimes H_2) X_i (H_1\otimes H_2) \Lambda_1(S^{\dagger}) =
Z_i$ for $i = 1,2$.  $|\psi_+\rangle$ is an entangled state, and both of
$M_1$ and $M_2$ are two-qubit operators. But the following state
\begin{eqnarray}
	|\psi\rangle 
	&=& \frac{1}{\sqrt{2}}( |\psi_+\rangle+Q_1 |\psi_+\rangle)
\\	&=& \frac{1}{\sqrt{2}}|0\rangle(|0\rangle+|1\rangle)
\
\end{eqnarray}
is a product of single qubit states and has stabilizers $Z_1$ and
$X_2$. Thus, we can first prepare $|\psi\rangle$ fault-tolerantly by
measuring $Z_1$ and $X_2$, and then measure $M_1$ alone to get the state
$|\psi_+\rangle$.  Equivalently, we can also first prepare the state
$\frac{1}{\sqrt{2}} (|\psi_+\rangle+Q_2|\psi_+\rangle) =
\frac{1}{\sqrt{2}}(|0\rangle+|1\rangle) |0\rangle$, which has stabilizers
$X_2$ and $Z_1$, and measure $M_2$ to obtain the state $|\psi_+\rangle$.
%
\subsubsection{Fault-tolerant preparation of the required ancilla for the
Toffoli gate}
The required ancilla for the  Toffoli gate
construction is 
\begin{eqnarray}
	|\psi_+\rangle 
	&=& U(H_1\otimes H_2) |000\rangle
\\	&=& \frac{1}{2}(|000\rangle+|010\rangle+|100\rangle+|111\rangle)
\,,
\end{eqnarray}
where $U$ is the Toffoli gate. The stabilizer of this state is
$M_i = U(H_1\otimes H_2) Z_i (H_1\otimes H_2) U^{\dagger}$ for $i = 1,2,$
and $3$. Using Eqs.~(\ref{eq:tofcom1})-(\ref{eq:tofcom2}), 
\begin{eqnarray}
	M_1 &=& X_1 \otimes {\sc cnot}_{23}
\,,
\\	M_2 &=& X_2\otimes {\sc cnot}_{13}
\,,
\\	M_3 &=& Z_3\otimes {\sc cz}_{12}
\,,
\end{eqnarray}
where {\sc cz} represents a controlled-$Z$, and the {\em ordered}
subscripts for {\sc cnot} and {\sc cz} specifies the control and target
bits. The operator that anticommutes with $M_i$ is $Q_i = U(H_1\otimes H_2)
X_i (H_1\otimes H_2) U^{\dagger}$, or $Z_1, Z_2$ and $X_3$ for $i = 1,2,$
and $3$, respectively.  Again, each of $M_i$ is a two-qubit operator, but
the following state
\begin{equation}
	|\psi\rangle = \frac{1}{\sqrt{2}}(|\psi_+\rangle+Q_1|\psi_+\rangle)
	= \frac{1}{\sqrt{2}}|0\rangle(|0\rangle+|1\rangle) |0\rangle
\end{equation}
can be prepared easily by measuring its stabilizers $Z_1,X_2$ and $Z_3$.
Then we only need to measure a single two-qubit operator $M_1$ on
$|\psi\rangle$ to obtain $|\psi_+\rangle$.
Equivalently, we can also first prepare the state $|\psi\rangle =
\frac{1}{\sqrt{2}}(I+Q_2) |\psi_+\rangle$ with stabilizers $X_1,Z_2$ and
$Z_3$ or the state $|\psi\rangle = \frac{1}{\sqrt{2}}(I+Q_3)
|\psi_+\rangle$ with stabilizers $X_1,X_2$ and $X_3$, and measure the
corresponding single stabilizer to obtain $|\psi_+\rangle$.

\subsection{Recursive preparation}
In this subsection, we will prove the following Theorem, which is used in
Sec.~\ref{sec:recursive}:\\ 
{\bf Theorem 2:} The encoded state $|\psi\rangle = UH^{\otimes
n}(|0\rangle^{\otimes n})$ for $U\in F_k$ can be prepared fault-tolerantly
by recursive application of one-bit teleportation.

First we have the following Lemma, which is a generalization of Lemma 3.\\
{\bf Lemma 4:}  $|\psi\rangle$ can be created fault-tolerantly if given 
$\mbox{St}(|\psi\rangle) = \{M_1,\ldots,M_n\}$, $\forall i,j$\\
(1) the cat-state-controlled-$M_i$ can be performed fault-tolerantly;\\
(2) there exists $Q_i$ such that $Q_i$ can be  performed
fault-tolerantly, $\{M_i,Q_i\} = 0$, and for $i\neq j, [M_i,Q_j] = 0$. \\
{\em Proof:} Since $M_i^2 = I$, by applying the cat-state-controlled-$M_i$
and measuring the cat state fault-tolerantly as before, we can project any
encoded state onto $\pm 1$ eigenstate of $M_i$. Then apply $Q_i$ if a $-1$
eigenstate is obtained. Using the same argument as in the proof of Lemma 3,
we can fault-tolerantly prepare the state $|\psi\rangle$. \\
{\bf Lemma 5:} If operations in $F_{k-1}$ and cat-state-controlled-$V$ for
any $V\in F_{k-2}$ can be performed fault-tolerantly using the one-bit
teleportation scheme, then $UH^{\otimes n}(|0\rangle^{\otimes n})$ for $U\in
F_k$ can be created fault-tolerantly.\\ 
{\em Proof:}
The stabilizers of $|\psi\rangle = UH^{\otimes n}(|0\rangle^{\otimes n})$ are
$M_i = UH^{\otimes n}Z_i (UH^{\otimes n})^{\dagger} = UX_iU^{\dagger} = U_x
X_i$ for some $U_x\in F_{k-1}$.  Define $Q_i \equiv UZ_iU^{\dagger} = 
Z_i$. $Q_i$ satisfies condition (2) of Lemma 4.

Since $M_i = U_x X_i$, the cat-state-controlled-$M_i$ is the product of
cat-state-controlled-$U_x$ and cat-state-controlled-$X_i$.  The
cat-state-controlled-$X_i$ is easily performed fault-tolerantly. Thus it
remains to show how to perform the cat-state-controlled-$U_x$ for $U_x\in
F_{k-1}$ fault-tolerantly.

As $U_x\in F_{k-1}$ is constructed with one-bit teleportation scheme using
the circuit of Eq.~(\ref{eq:basic2}), where $A_i = H, B_i = I$ and $D_i =
X_i$, to perform cat-state-controlled-$U_x$, we need to perform
cat-state-controlled-$E$, cat-state-controlled-$U_xX_iU_x^{\dagger}$, and
to prepare the ancilla $U_xH^{\otimes n}(|0\rangle^{\otimes n})$
fault-tolerantly.  As $E\in C_2$ and $U_x X_i U_x^{\dagger} = U_x'X_i$ with
$U_x' \in F_{k-2}$, both of cat-state-controlled-$E$ and
cat-state-controlled-$U_xD_iU_x^{\dagger}$ can be performed
fault-tolerantly. Next, the state $U_xH^{\otimes n}|0\rangle^{\otimes n}$
has stabilizers $M_i' = U_xX_iU_x^{\dagger} = U_x' X_i$ with $U_x'\in
F_{k-2}$, which satisfies both conditions of Lemma 4 and can therefore be
prepared fault-tolerantly. Thus cat-state-controlled-$U_x$ can be performed
fault-tolerantly.  This completes the proof of Lemma 5.

In fact, what we have shown in the proof of Lemma 5 is that if operations
in $F_{k-1}$ and the cat-state-controlled-$V$ for $V\in F_{k-2}$ can be
performed fault-tolerantly, then the cat-state-controlled-$U$ for $U\in
F_{k-1}$ and operations in $F_k$ can be performed fault-tolerantly.  This
is because according to Sec.~\ref{sec:recursive}, fault-tolerant
construction of $F_k$ gates only require fault-tolerant $F_{k-1}$ gates and
an ancilla $UH^{\otimes n}(|0\rangle^{\otimes n})$ for $U\in F_{k}$.

Since both the operations in $F_2$ and the cat-state-controlled-$U$ for
$U\in F_1$ can be performed fault-tolerantly, by induction, operations in
$F_k$ and the cat-state-controlled-$U$ for $U\in F_{k-1}$ can be
constructed fault-tolerantly, with which we can fault-tolerantly prepare
the encoded state $UH^{\otimes n}(|0\rangle^{\otimes n})$ for $U\in F_k$.


\section{Alternative remote {\sc cnot} circuit}
\label{sec:gottesmancn}
In this section, we re-derive the remote {\sc cnot} construction, given
in~\cite{Gottesman98}, using one-bit teleportation.  A remote {\sc cnot}
between the states $|\alpha\rangle$ and $|\beta\rangle$ belonging to Alice
and Bob, respectively, can be performed by a four step procedure:
(1) Alice swaps her state $|\alpha\rangle$ with an ancilla $|0\rangle$, (2)
Alice sends the teleported $|\alpha\rangle$ to Bob using $X$-teleportation,
(3) Bob applies {\sc cnot} from $|\alpha\rangle$ to $|\beta\rangle$, and
(4) Bob teleports $|\alpha\rangle$ back to Alice using $Z$-teleportation.
Steps (2) and (4) involve prohibited operations. 
Here is a circuit representation: 
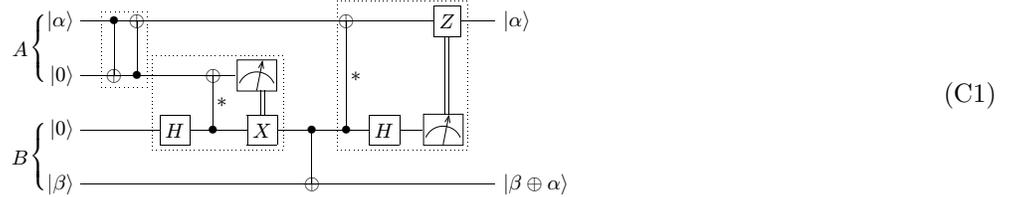
\begin{equation}  
\parbox{3.25in}{\psfig{file=xor1.epsf,width=3.0in}}
\end{equation}
The two prohibited {\sc cnot}s are labelled with asterisks. 
They can be commuted backwards to obtain the equivalent circuit: 
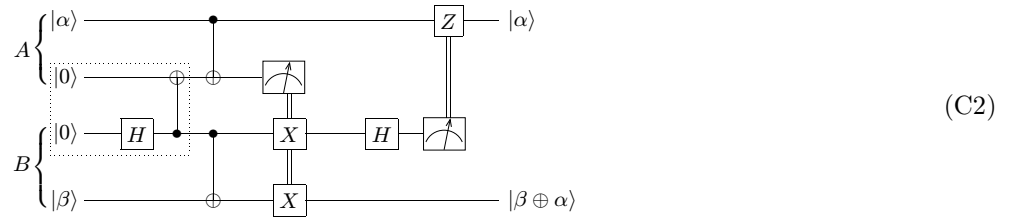
\begin{equation}
\parbox{3.25in}{\psfig{file=xor4.epsf,width=3.0in}}
\end{equation}
which again reduces prohibited operations to some specific shared 
entangled state.  
%


\begin{references}
\bibitem{notation} We write $UEU^{\dagger}$ instead of 
$(U \otimes I^{\otimes n}) E (U^{\dagger}\otimes I^{\otimes n})$
for simplicity.  Unimportant identity operators are suppressed 
throughout the paper.
\bibitem{coincidence} Note that the $n$ operators $UD_iU^\dagger$ to be
performed conditioned on the measurement outcomes are precisely the $n$
stabilizers of the ancilla state $UA(|0\rangle^{\otimes n})$.  This coincidence
is due to the structure of the teleportation circuits.  See also the
examples in Sec.~\ref{sec:3examples}.
\end{references}
\end{document}